\documentclass[a4,12pt]{article}
\usepackage{amsmath}
\usepackage{epsfig}
\def\sumint{\mathop{\rlap{$\sum\null$}\mkern-2mu\int}\nolimits}
\begin{document}
\begin{center}
\Huge{Fermion Number Fractionization}
\end{center}
\vspace{1cm}
\begin{center}
{\bf Kumar Rao, Narendra Sahu and Prasanta K. Panigrahi}\\
{\em Theoretical physics division, Physical Research Laboratory, 
Ahmedabad, 380 009, India}
\end{center}
\vskip1cm                                                             

{\bf Solitons emerge as non-perturbative solutions of non-linear wave equations in classical and quantun theories. These are non-dispersive and localised packets of energy-remarkable properties for solutions of non-linear differential equations. In presence of such objects, the solutions of Dirac equation lead to the curious phenomenon of `fractional fermion number', a number which under normal conditions takes strictly integral values. In this article, we describe this accidental discovery and its manifestation in polyacetylene chains, which has led to the development of organic conductors.}


\section*{Introduction}
 The Nobel prize in chemistry for the year 2000 was awarded to 
Alan J. Heeger, Alan G. MacDiarmid and Hideki Shirakawa for their 
discovery and development of conducting polymers. Although the Nobel 
prize was awarded in chemistry, the theoretical explanation of 
conducting polymers came from a completely unexpected area: Quantum field 
theory. In 1976 Roman Jackiw and Claudio Rebbi showed that when fermions\footnote{Particles carrying half integral spins like electrons, protons, neutrons etc. are known as fermions, whereas particles with integral spin, like photons, pions are known as bosons.}
interact with background fields with a nontrivial topology, like a soliton, the ground state of the fermion-soliton system can have {\it 
fractional} fermion numbers of $\pm \frac{1}{2}$ ! More interestingly, their work demonstrated how 
fermion-soliton interactions in the polyacetylene molecule can induce such novel quantum 
numbers in that system. {\it Jackiw was awarded the Dirac medal of the 
International Center for Theoretical Physics (ICTP) in 1998 for his many 
contributions to field theory, including the discovery of fractional 
fermion number} \cite{jackiw}. In the following, we describe the broad ideas behind fractionization of the fermion number and its manifestation in different 
physical systems. For a rigorous introduction to fermion number fractionization with many applications, but at an advanced level, see the review \cite{niemi}.

\section{The Dirac sea}
The marriage of quantum mechanics with special relativity gave rise to 
a number of successes and problems. To start with, for free particles the Schr$\ddot{o}$dinger wave equation, based on the non-relativistic dispersion relation $E=\frac{{\bf p}^2}{2m}$ needed to be modified. The Klein-Gordon (KG) equation, which was a wave equation based on the 
relativistic energy-momentum relation $E^2={\bf p}^2 c^2 +m^2 c^4$, did not 
correctly describe the spectrum of the Hydrogen atom. More importantly the 
interpretation of $\Psi^{\dagger} \Psi$ as a probability density could not 
hold, since it could assume negative values. This problem was traced to the fact that the KG 
equation is second order in time. To rectify this, in 1928 Dirac constructed 
a differential equation which was first order in space {\it and} time, as 
demanded by relativistic covariance. Though the (multicomponent) Dirac 
equation correctly accounted for the half-integral spin of the electron, it 
was plagued by the existence of an infinite number of unphysical negative 
energy solutions, as was also the case with the KG equation. {\it It is worth emphasizing that one cannot just ``discard'' the negative energy solutions since both positive and negative energy states are needed to describe a  complete set of states, as should be the case for a Hermitian Hamiltonian}. An electron in 
a positive energy state could, under a perturbation, cascade to states with progressively lower energy, in the process emitting radiation with 
a continuous spectrum. Hence, atoms would collapse; the Dirac equation could 
not describe the dyanamics of a stable fermionic system. To avoid this fate 
of the electrons in atoms, Dirac postulated that all the negative energy 
states be filled with electrons. The Pauli exclusion principle would then 
prevent electrons from positive energy states from occupying negative energy 
states. The infinite `sea' of negative energy electrons, with their infinite 
charge and mass, called the Dirac sea, was regarded to be the vacuum, i.e., 
the ground state of the system. The infinite charge and mass of the negative 
energy electrons was postulated to be unobservable; only a ``departure'' 
from the vacuum state, like an electron in one of the positive energy states, 
was supposed to be observable. If one of the negative energy electrons could 
gain enough energy (say by an electric field) to cross the energy gap, it 
would appear as a real electron. The vacancy left behind in the negative 
energy state has all the properties of a ``positive electron'' - for example, it would move in the opposite direction, as compared to a negative electron, 
under the influence of an electric field. The ``positive electron'', called 
a positron, was discoverd in 1932, brilliantly confirming Dirac's hypothesis.

The marriage of special relativity and quantum mechanics inherently describes a {\it multiparticle theory} - a single 
particle interpretation of the Dirac equation would not be consistent. 
This led to the birth of relativistic quantum field theory, where the one particle wave function $\Psi$ 
is interpreted as an {\it operator} acting on the Hilbert space of quantum 
states (number states or Fock states which contain a given number of particles with fixed momenta), which obeys the appropriate equation 
of motion. It was found that spin zero particles obeyed the KG equation and spin half particles, the Dirac equation. $\Psi$ is expanded in terms of plane wave solutions 
of the appropriate equation of motion, with the expansion 
coefficients being ``annihilation'' ($a_{\textrm{\bf p}}$) and ``creation''  
($a_{\textrm{\bf p}}^\dagger$) operators acting on the number states of 
the Hilbert space. For example, the KG field operator is written as
\begin{equation}
\Psi (\textrm{\bf x})=\int \frac{d^3 p}{(2 \pi)^3} \frac{1}{\sqrt{2 
E_{\textrm{\bf p}}}} \,\,\left(a_{\textrm{\bf p}}e^{i \textrm{\bf p} 
\cdot \textrm{\bf x}} +a_{\textrm{\bf p}}^{\dagger}e^{-i \textrm{\bf p} 
\cdot \textrm{\bf x}}\right),
\label{Psi}
\end{equation}
where the integral is over all the plane wave solutions $e^{\pm i 
\textrm{\bf p} \cdot \textrm{\bf x}}$ of the KG equation, labelled by the 
momentum {\bf p}. $E_{\textrm{\bf p}}=+\sqrt{{\bf p}^2 c^2 +m^2 c^4}$ 
is the relativistic energy of the plane wave state.The factor of $\sqrt{2 
E_{\textrm{\bf p}}}$ is required to make the norm of the one particle 
states Lorentz invariant and is the standard normalization used in 
relativistic field theory. One quantizes the system by imposing the 
usual commutation relation between the (generalised) field coordinate 
and its canonical momentum $\Pi$:
\begin{equation}
\left[\Psi (\textrm{\bf x}), \Pi (\textrm{\bf y})\right]=i\,\delta ^{(3)}
(\textrm{\bf x - y}).
\label{commf}
\end{equation}
$\Pi$ is the usual canonical momentum obtained from the Lagrangian density 
by differentiating with respect to the time derivative of $\Psi$; this time 
dependence is not explicitly shown in (\ref{Psi}). In terms of the creation 
and annihilation operators this commutation relation reads
\begin{equation}
\left[a_{\textrm{\bf p}}, a_{\textrm{\bf k}}^{\dagger}\right]=(2 \pi)^3 
\delta^{(3)}(\textrm{\bf p - k}) \qquad (\textrm{Spin zero field}).
\label{comma}
\end{equation}
Note that these commutation relations diverge if they are evaluated at the 
same space-time point or the same momentum mode. Particle and antiparticle 
states are interpreted as excitations of the vacuum by suitable creation 
operators acting on the vacuum state $|0\rangle$ (which is a zero particle 
state) and have positive energy:
\begin{equation}
|\textrm{\bf p}\rangle \sim  a_{\textrm{\bf p}}^{\dagger}|0\rangle .
\end{equation}
The antiparticle state is created with the creation operator $ b_{\textrm{\bf p}}^{\dagger}$. 

For the Dirac field an eigenmode expansion similar to (\ref{Psi}) holds, but 
with eigenfunctions of the corresponding Dirac equation. One crucial 
difference between the KG and Dirac field operators is that, while the 
former obey the commutation relation (\ref{commf}), the latter obeys 
an {\it anticommutation} relation. This also holds true for the 
corresponding creations and annihilation operators:
\begin{equation}
\{a_{\textrm{\bf p}},a_{\textrm{\bf k}}^{\dagger}\} \equiv a_{\textrm{\bf p}} 
a_{\textrm{\bf k}}^{\dagger}+ a_{\textrm{\bf k}}^{\dagger}  
a_{\textrm{\bf p}}=(2 \pi)^3 \delta^{(3)}(\textrm{\bf p - k}) 
\qquad (\textrm{Spin half field}).
\end{equation}
It is this anticommutation relation which implements the observed fact that 
spin $\frac{1}{2}$ fields (and generally all fields with half integer spin) 
obey Fermi statistics (which states that two fermions cannot occupy the same 
state or mode).

Since $\Psi$ is no longer a wave function, but an operator, it is not 
necessary that its norm be positive definite. Interpreting the vacuum of 
a fermionic system as an infinite sea of filled electrons did not seem 
to be necessary in this framework.

However, as we will show, the Dirac sea does have real and observable 
effects. The reader may be aware of two other manifestations of the 
vacuum-that of the Casimir effect and the Klein paradox. In this article, 
we outline the relatively less well known phenomenon of fractional fermion 
number, which can be interpreted as a result of nontrivial changes in the 
structure of the Dirac sea. While the same results can be more rigorously 
derived using the formalism of quantum field theory, the Dirac sea picture 
has the advantage of making a connection with nonrelativistic quantum 
mechanics, where the wave function $\Psi ^{\dagger} \Psi$ is interpreted 
as a probablity or number density of particles.

\section{The Fermion Number}
If a Lagrangian with Dirac fermions is invariant under a global $U(1)$ phase 
transformation ($\Psi \to e^{i \alpha}\Psi$), by Noether's theorem there exists 
a conserved current associated with it. The corresponding conserved charge, called the {\it fermion number} of the system, is
\begin{equation}
N=\int d^{3}\textrm{\bf x} \,\Psi^{\dagger}(\textrm{\bf x}) \Psi(\textrm{\bf x}).
\end{equation}
In terms of the creation and annihilation operators this is written as
\begin{equation}
N=\int \frac{d^3 p}{(2 \pi)^3} \sum _{s} \left( a_{\textrm{\bf p}}^{s \dagger} 
a_{\textrm{\bf p}}^{s} + b_{\textrm{\bf -p}}^{s} b_{\textrm{\bf -p}}^{s \dagger}\right),
\end{equation}
where $a_{\textrm{\bf p}}^{s \dagger}$ and  $b_{\textrm{\bf p}}^{s \dagger}$ 
are the creation operators of the fermion and antifermion respectively and 
$s=1,2$ are the two spin states of the fermion/antifermion. Because of the 
anticommutation relations of the $b$ and $b^\dagger$ (similar to (\ref{comma})), 
we get an infinite contribution when the $b$ is taken to the right of $b^\dagger$, 
since both are evaluated at the same momentum mode. This infinite contribution 
to the fermion number is simply the contribution of the Dirac sea. To properly 
define the fermion number we have to {\it subtract} the contribution of the 
vacuum state; this can simply be achieved by the following commutator \cite{niemi}:
\begin{equation}
N=\frac{1}{2} \int d^3 \textrm{\bf x}\left[\Psi ^{\dagger}, \Psi \right].
\label{num}
\end{equation}
This is the {\it regularised} or {\it normal ordered} form of the number operator.

By introducing a suitable regularization for the operators to avoid 
dealing with infinities in the form of delta functions, a careful 
calculation shows that the number operator (\ref{num}) becomes
\begin{equation}
N=\int \frac{d^3 p}{(2 \pi)^3} \sum _{s} \left( a_{\textrm{\bf p}}^{s \dagger} 
a_{\textrm{\bf p}}^{s} - b_{\textrm{\bf p}}^{s \dagger} b_{\textrm{\bf p}}^{s}
\right)  -\frac{1}{2}\eta ,
\label{num2}
\end{equation}
where the c-number piece $(\eta/2)$ is the {\it spectral asymmetry} of the 
system, i.e., the {\it difference between the number of positive and negative 
energy states of the fermion spectrum.} This assumes that the fermion 
spectrum has no zero mode, i.e., a state with zero energy, otherwise the 
formula is general and in particular gives the fermion number when fermions 
interact with background fields. We will consider the case of the zero mode 
solution later. The operators $a_{\textrm{\bf p}}$ and  $b_{\textrm{\bf p}}$ 
annihilate the vacuum state, and hence the fermion number of the vacuum is 
simply,
\begin{equation}
\langle N\rangle _0=\langle 0|N|0\rangle=-\frac{1}{2}\eta .
\end{equation} 

\begin{figure}
\epsfig{file=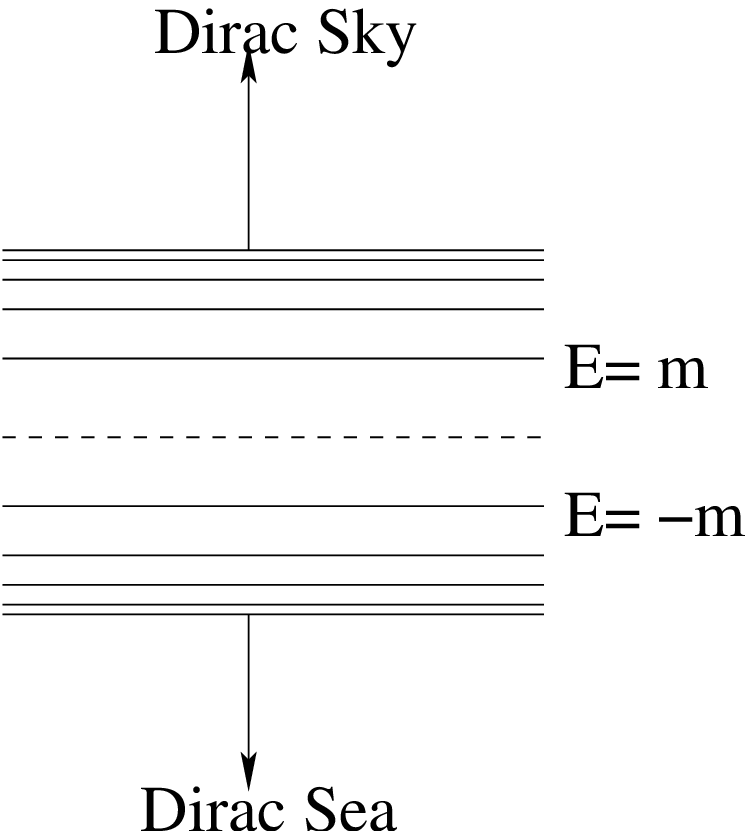, width=0.4\textwidth}
\epsfig{file=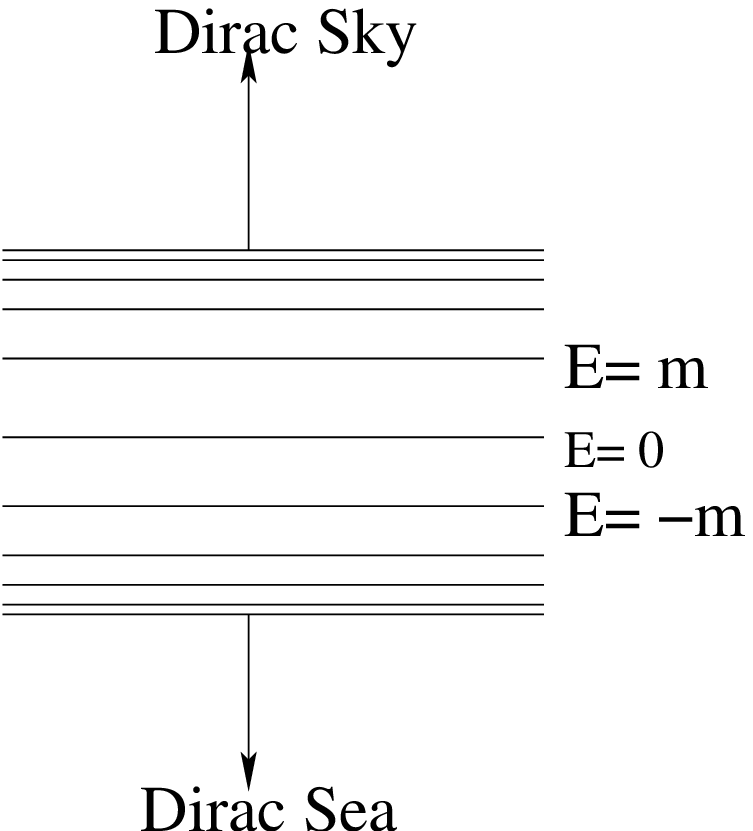, width=0.4\textwidth}
\caption{The energy spectrum of free fermions of mass $m$ (left) and in the presence of the soliton (right). We notice that there is a zero energy bound state in between the mass gap for the case of the soliton.} 
\end{figure}

For free fermions with no interactions, the fermion spectrum is exactly 
symmetric between the positive and negative energy states and invariant 
under charge conjugation. Hence the spectral asymmetry is zero and so 
is the fermion number, as expected. The operator pieces in (\ref{num2}) 
always have integral eigenvalues, with the states created by 
$ a_{\textrm{\bf p}}^{s \dagger}$ having fermion number +1 and those created 
by  $ b_{\textrm{\bf p}}^{s \dagger}$ having fermion number -1. We now give a 
heuristic argument, based on the Dirac sea picture, of the origin of the 
$-\frac{1}{2}\eta$ term. Consider fermions interacting with a background 
field, which is described classically. A complete analysis taking into 
account quantum fluctuations of the background field is a more 
difficult problem, which we do not address in this article. The approximation 
of classical fields interacting with second quantized fermions is 
called ``semiclassical''. The corresponding eigenfunctions and energy 
eigenvalues are obtained by solving the one particle Dirac equation with 
the background potential. We denote the energy eigenfunction as $\Psi _E$ in 
the presence of the field and $\phi _E$ for the free Dirac field, in which 
case the $\phi _E$ are just plane waves. Note that both $\Psi _E$ and 
$\phi _E$ are spinors as they are solutions of the matrix Dirac equation). 
The background field would distort the fermion spectrum as compared to the 
free case and, in general, there could also be {\it bound states} in between 
the mass gap. As mentioned before, we define the regularised fermion number 
by subtracting the contribution of the free Dirac vacuum; the fermion 
number is measured relative to the free vacuum (which, we recall, consists 
of all the negative energy states being occupied in the Dirac sea 
picture). Interpreting $|\Psi|^2$ as a number density, as in 
non-relativistic quantum mechanics, the vacuum fermion number density of 
the interacting fermion system is given by
\begin{eqnarray}
\rho (\textrm{\bf x})&=&\sumint _{-\infty}^{0}d E |\Psi _E|^2 - \int _{-\infty}^0 dE 
|\phi _E|^2 \\ \nonumber
&=& \sumint _{-\infty}^{0}d E |\Psi _E|^2 -\frac{1}{2}  \int _{-\infty}^{\infty} 
dE |\phi _E|^2 \\ \nonumber
&=&\sumint _{-\infty}^{0}d E |\Psi _E|^2 -\frac{1}{2}\sumint _{-\infty}^{\infty}
d E |\Psi _E|^2 \\ \nonumber
&=&-\frac{1}{2}\sumint _{0}^ {\infty}d E |\Psi _E|^2 +\frac{1}{2}\sumint _{-\infty}^{0}
d E |\Psi _E|^2 
\end{eqnarray}
The integral/sum is over all the states with negative energy, with an integral over the continuum states and summation over the bound states, if any in the case of $\Psi _E$. Of course, the free Dirac field $\phi_E$ has no bound states and thus no summation is required. The second step follows because the free Dirac spectrum is symmetric with respect to the energy and thus an integral over the negative energy states 
is simply one half the integral over all states. The third step is because of 
the completeness relation-the total number of states is the same regardless of 
the presence or absence of an interaction. Hence, assuming the states are 
normalized to unity, the vacuum fermion number is
\begin{eqnarray}
\langle N\rangle _0&=&\int d^3 \textrm{\bf x} \, \rho (\textrm{\bf x})\\ \nonumber
&=&-\frac{1}{2} \left[\sumint_0^{\infty}dE -\sumint_{-\infty}^{0}dE\right]
\end{eqnarray}
The integral above {\it counts the number of states} within the specified energy interval; 
the quantity in brackets is thus the spectral density $\eta$. If $\eta \neq 0$, the 
fermion number of the vacuum is {\it not} zero, and, in general, it need not even be 
a rational number.

\subsection{Zero Modes}
An interesting situation arises when a fermion soliton system has a zero energy solution. 
For simplicity consider fermions interacting with a classical background scalar field with a double well 
potential structure in one spatial dimension. The potential energy density of the scalar field can be taken in the convenient form $V(\phi)=\frac{1}{2}(\phi^2-v^2)^2$. The energy of the $\phi$ field is minimised when 
$\phi=\pm v$, i.e, the ground state of the system (when the scalar field is quantized) is 
doubly degenerate. Once the ground state is chosen, the $\phi \leftrightarrow -\phi$ symmetry is 
broken. The resulting Dirac equation describes a free massive fermion and the resulting 
spectrum has a mass gap of $2m$, $m$ being the mass of the fermion. There is no zero 
energy mode. This is called the vacuum sector. The eigenmode expansion of the field 
operator gives
\begin{equation}
\Psi(\textrm{\bf x})=\int \frac{dk}{2 \pi} \left[b_k u_k (\textrm{\bf x})+d_k^{\dagger} 
v_k(\textrm{\bf x})\right]
\end{equation}
where $k$ labels the states and $u_k$ and $v_k$ are the positive and negative energy 
solutions of the massive Dirac equation (which are just plane waves). From (\ref{num}) 
the fermion number operator is then
\begin{equation}
N=\int \frac{dk}{2 \pi} \left[b_k^{\dagger}b_k -d_k^{\dagger} d_k\right] .
\end{equation}
As we have seen before $N$ has only integral eigenvalues.

Whenever we have degenerate vacua, as in the double well case, there also exist solutions 
of the scalar field, called {\it solitons} which interpolate between the two vacua, i.e 
between -$v$ to $v$. Perturbations around the vacua at $\pm v$ would never reveal the 
presence of the soliton, which are thus inherently non-perturbative objects. Whenever 
fermions interact with such solitons, the corresponding Dirac equation has a zero 
energy solution in the middle of the mass gap, a nontrivial result which can be proved 
generally. This is the soliton sector. For a lucid introduction to solitons and their many applications, including fermion fractionization, see \cite{rajaraman}. To compute the fermion number in this sector the 
exact form of the zero energy eigenmode is not required. We have to take into account this 
extra eigenmode solution when expandng $\Psi$ in a complete set of states:
\begin{equation}
\Psi_S(\textrm{\bf x})=\sumint_{k\neq 0}\frac{dk}{2 \pi}\left[b_k\tilde{u_k}(\textrm{\bf x}) + 
d_k^{\dagger}\tilde{v_k}(\textrm{\bf x})\right]+ a\, \eta _0 (\textrm{\bf x}) .
\end{equation}
Here $\eta_0$ is the zero energy eigenfunction and $a$ is the corresponding expansion 
coefficient. The state $\eta_0$ is charge self conjugate, i.e, under the charge conjugation 
operator it goes into itself. The $k\neq0$ constraint means that the integral/sum is over 
all the non-zero energy states. Substituting this in (\ref{num}) one can show that the 
fermion number operator is
\begin{equation}
N_S=\sumint _{k\neq 0}\frac{dk}{2 \pi} \left[b_k^{\dagger}b_k -d_k ^{\dagger}d_k \right] + 
a^{\dagger} a - \frac{1}{2} .
\label{sol}
\end{equation}
The term $a^{\dagger} a$ counts the number of electrons in the zero energy state. Since 
the state has zero energy, the energy of the system does not change if it is occupied 
by a fermion or not. Hence the ground state is doubly degenerate. If the state $\eta_0$ 
is not occupied the fermion number is $-\frac{1}{2}$, corresponding to the c-number 
piece in (\ref{sol}) above. If $\eta_0$ is occupied, then the fermion number is 
$-\frac{1}{2}+1=+\frac{1}{2}$. We emphasize that the fermion number of $\pm\frac{1}{2}$ 
for the two degenerate vacua is an {\it eigenvalue} of the number operator, not just 
an {\it expectation value}. Expectation values can, in general, be fractional in 
quantum mechanics. For example, a particle tunnelling back and forth in a double well 
potential spends half its time in either well; the probablity of finding it in 
a given well is one half.

Though the eigenvalue of $\pm \frac{1}{2}$ is surprising, it is even more amazing that 
this relativistic field theory model of fermion-soliton interaction finds an application 
in a ``down to earth'' condensed matter system-the polyacetylene molecule! In the 
following we outline this interesting phenomenon.

\section*{Solitonic excitations in polyacetylene}
\begin{figure}
\centerline{\epsfig{file=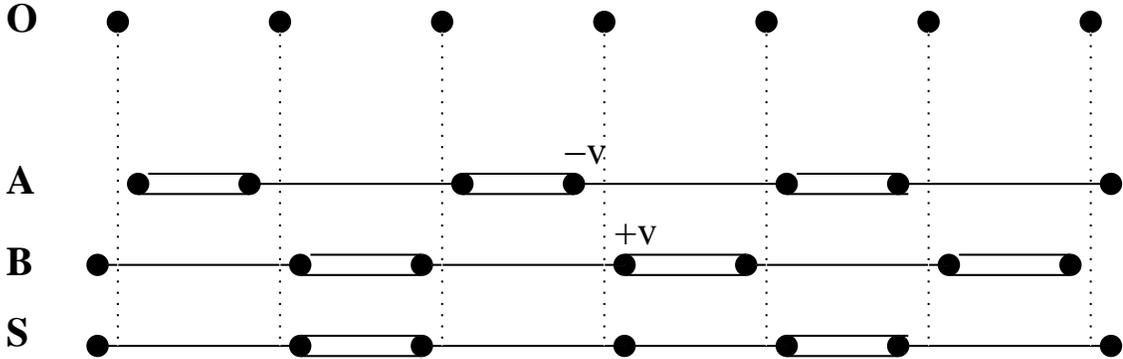,width=1.0\textwidth}}
\caption{The schematic representation of the unstable carbon chain O, 
the two degenerate ground states A and B, and the soliton state S in 
polyacetylene chain.}
\label{single_soliton}
\end{figure}
We have seen that in order to get fractional fermion numbers, there must 
be a distortion in the Dirac spectrum ($\eta \neq 0$), or zero modes or an 
asymmetrical number of bound states must be present. As we have seen, this can happen 
when fermions interact with topological objects like solitons.

All the ingredients necessary for fermion number fractionization come 
together in polyacetylene \cite{su}. This is a one dimensional array of 
carbon atoms which can form one of the two degenerate ground states: 
A ($\cdots 212121 \cdots$) and B ($\cdots 121212 \cdots$) as shown in 
figure (\ref{single_soliton}), where `1' and `2' are the schematic 
representation of single and double bonds in the configuration of A and 
B. The degeneracy arises from a spontaneous breaking of the left-right 
symmetry in the one dimensional chain O, and manifests itself in an 
alteration of the bonding pattern, as illustrated in the two ground states 
A and B. The left-right symmetric state with respect to any carbon atom in a 
long chain carbon atoms is very much unstable. This instability of the 
carbon atoms is called `Pierels instability'. It is a generic tendency of 
all systems to remain in a state having minimum energy configuration. Therefore, 
the system slips into any one of the degenerate ground state A or B. These 
states are called conjugated chains of polyacetylene having
alternate single and double bonds. These states are energetically stable with a 
large band gap of order $1.5 eV$ and thus behave as semiconductors. However, 
polyacetylene can be transformed to a conductor by doping it with either an electron 
donor or an electron acceptor. This is reminiscent of doping of silicon based 
semiconductors where silicon is doped with either arsenic or boron. However, 
while the doping of silicon produces a donor energy level close to the valence 
band, this is not the case with polyacetylene. In the latter case the doping leads 
to the formation of a defect called soliton (a non-trivial twist of the conjugated 
chain) in the alternate bonding pattern of carbon atoms as shown in the third chain 
S ($\cdots 121121 \cdots$) of figure (\ref{single_soliton}). This results 
in the creation of a new localized electronic state in the middle of the energy gap. 
At high doping levels, the charged solitons interact with each other to form a 
soliton band which can eventually merge with the band edge to create true metallic 
conductivity and thus polyacetylene behaves as a `conducting polymer'.

\section*{Charge fractionization in polyacetylene}
\begin{figure}
\centerline{\psfig{file=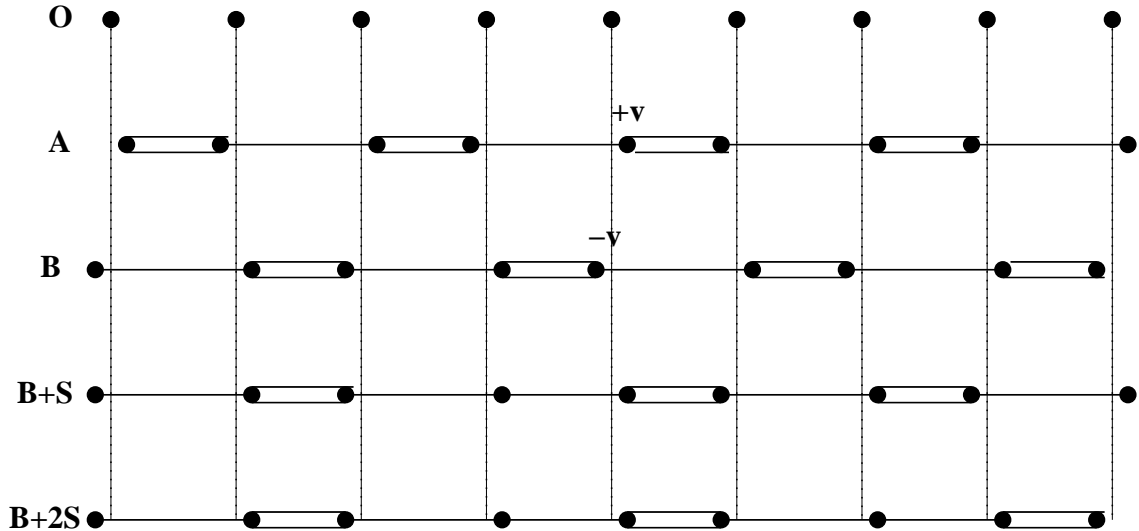,width=1.0\textwidth}}
\caption{The schematic representation of the unstable carbon 
chain O, the two degenerate ground states A and B, the soliton 
states B+S and B+2S in polyacetylene chain.}
\label{double_soliton}
\end{figure}

Let us consider a defect obtained by removing a bond from 
the conjugated chain B ($\cdots 12121212 \cdots$) at the 
fourth link in the form of $\cdots 12111212 \cdots$. By 
exchanging bonds between fifth and sixth links we arrive 
at $\cdots 12112112\cdots $. This is nothing but the 
$B+2S$ state shown in figure (\ref{double_soliton}). This 
displays the original defect (i.e., B+2S) as two more 
elementary ones. Indeed the elementary defect B+S 
($\cdots 12112121\cdots$) if continued without further 
disruption of the order, is a domain wall interpolating 
between the ground state B ($\cdots 12121212 \cdots$) on the 
left and A ($\cdots 212121 \cdots$) on the right. The fact that 
by removing one bond we produced two domain walls, strongly 
suggests that each domain wall would carry a fractional charge 
$e/2$ and spin $\pm 1/4$. In reality bonds represent pairs of 
electrons (resulting in a net charge 2e) with opposite spin 
(resulting in a net spin 0) and so we don't get any fractional 
charge. But we do find something quite unusual, a domain wall 
acquires a charge $e$ with spin zero. Charge and spin which 
normally occur together have been separated, showing an indirect 
signature of charge fractionization in polyacetylene chain.    

The necessary ingredients for fermion number fractionization 
in polymer systems are: (1) the phonon fields, the lattice 
vibrations, which measure the displacement, and (2) left and right moving 
electrons. 
The valance electrons resemble the negative energy filled states, while 
the conduction electrons populate the positive energy states. 
Although fractional fermion number was first realized in 
polyacetylene chain through solitonic excitation, it is no 
more confined to it. It is observed in many different 
areas in condensed matter physics. In particular, fractional quantum 
Hall effect and fractional charge in short noise experiments 
are prime among them. 

\section{Topological quantum numbers}
From classical/quantum mechanics and field theory, we know that whenever the 
Lagrangian of a system is invariant under a continuous transformation, there 
is a conserved current and charge corresponding to that transformation. This 
is Noethers theorem. In fact, as we mentioned in the beginning of section (2), 
the fermion number is the conserved charge corresponding to $U(1)$ phase 
invariance of the fermion field. 

The fermion number in the solitonic sector, apart from being a Noether charge, is 
also {\it topological}. By this we mean the following: It depends only on the 
asymptotic properties of the background field, i.e., its value is fixed at the boundaries 
of space and {\it does not depend on the detailed profile of the soliton in configuration 
space}. For example, for the potential $V(\phi)=\frac{1}{2}(\phi^2 -v^2)^2$ in one 
dimension, the soliton solution $\phi(\textrm{x})\sim \tanh$(x) goes from -$v$ 
to +$v$ as x goes from $-\infty$ to $+\infty$. If one changes the profile of the 
soliton, {\it keeping its asymptotic values $v$ fixed}, the fermion number 
does not change! In general, one would expect that the details of the fermion-soliton 
interaction will change as the profile of the soliton is changed; this involves 
solving the Dirac equation with a different potential, which changes the energy 
levels. Since we have related the fractional part of the fermion number to the 
spectral asymmetry, one would expect that the fermion number would also change. However, 
using very general and powerful mathematical techniques (like the famous 
Atiyah-Singer index theorem), it has been shown, in a large class of field 
theory models, that the fractional part of the fermion number is independent of 
changes in the profile of the background field. The solitons themselves are stable, 
i.e, cannot change continuously from one sector to another (or one vacua to 
another since each choice of vacua characterises the soliton) because it would 
require an infinite amount of energy to do so.

That the fractional part of the fermion number is topological, apart from being an 
important result in itself, is also crucial for many calculations. For example, very 
often a variational calculation is done to minimise the energy of the system, by 
varying the profile of the interacting fields. It is imporatnt that the quantum 
numbers characterising the system do not change during such a process. In the 
Skyrme model (a model in nuclear physics involving pion fields which admits soliton solutions in three dimensions) with fermions for instance, protons are interpreted as solitons 
or ``knots'' in the Skyrme background carrying a charge due to the spectral asymmetry 
induced by the Skyrme field. If that is the case, the charge of the proton should 
not change in a variational calculation involving the Skyrme field and that this is 
so is ensured by the topological nature of the induced fermion number. A lucid description of the Skyrme model and other soliton models of the nucleon can be found in \cite{rajat}.

\section{Summary}
In summary, we have shown, in a very simple setting, how the vacuum of a system of fermions can play an important part in generating novel phenomena like fractional fermion number. Exotic ideas derived from theoretical physics and mathematics may not be far removed from reality. The close interaction between theoreticians and experimentalists may lead to their physical realization on surprisingly down to earth systems as the above example of polyacetylene reveals. Many other related concepts like anyons, fractional statistics and spin are still to be confirmed by experiment. The recent discovery of new materials like graphene having a two dimensional structure and making of Bose-Einstein condenstates have opened the lower dimesional world for deeper exploration.

\end{document}